\documentclass{Rinton-P9x6}
\usepackage{amsmath,amssymb}
\newcommand{\dket}[1]{| \, #1 \rangle\!\rangle}
\newcommand{\dbra}[1]{\langle\!\langle #1 \, |}
\begin{document}
\title{Optimization of quantum universal detectors} 
\author{G. M.  D'Ariano, P. Perinotti, M. F. Sacchi} 
\address{QUIT Group, 
Unit\`a INFM and Dipartimento di Fisica ``A. Volta'', \\
 Universit\`a  di Pavia, via A. Bassi 6, I-27100 Pavia, Italy}
\maketitle

\abstracts{ The expectation value $\langle O\rangle$ of an arbitrary
  operator $O$ can be obtained via a universal measuring apparatus
  that is independent of $O$, by changing only the data-processing of
  the outcomes.  Such a ``universal detector'' performs a joint
  measurement on the system and on a suitable ancilla prepared in a
  fixed state, and is equivalent to a positive operator valued measure
  (POVM) for the system that is ``informationally
  complete''. The data processing functions generally are not unique,
  and we pose the problem of their optimization, providing some
  examples for covariant POVM's, in particular for $SU(d)$ covariance group.}

\par Universality and programmability are crucial features in quantum
technology, for communication, processing, and storage of information.
Different tasks should be achieved by a basic set of devices, that
would allow to perform different kinds of quantum information 
processing, such as in quantum computation\cite{pop,nie},
teleportation\cite{ben,bra}, entanglement detection\cite{wit}, and
entanglement distillation\cite{ben2}.  In particular, a universal
detector\cite{prl} achieves the estimation of the ensemble average of
an arbitrary operator by changing only the data processing of the
outcomes.  In some way it is analogous to a quantum tomographic apparatus
\cite{tomo}: however, the latter would typically require a {\em
  quorum} of observables---corresponding to a set of devices or to a
single tunable device---whereas a universal detector would measure
only a single fixed observable on an extended Hilbert space that
includes a suitable ancilla.

\par Universal detectors can be characterized via a necessary and
sufficient condition given in terms of "frames of operators"
(i. e. spanning sets of operators), and can be achieved via Bell
measurements, which are described by projectors on
maximally entangled states \cite{prl}. Entanglement, however, is not
an essential ingredient, and there are universal detectors which are
described by separable POVM's as well\cite{mauro}.

\par When attention is restricted to the system Hilbert space only,
universal detectors are equivalent to informationally complete
(shortly ``info-complete'') POVM's\cite{Busch}, which are frames made
of positive operators. Info-complete POVM's are necessarily
not-orthogonal, whence universal detectors have a more physical
counterpart, in terms of an observable and an apparatus ancilla. 

\par When using a universal detector the ensemble average of an
arbitrary operator is estimated by choosing the appropriate data
processing function of the measurement outcomes.  As we will see, the
data processing functions are generally not unique, and are related to
the concept of {\em dual operator frame}. In the following, after
reviewing the main results on universal detectors, we pose the problem
of optimization of data-processing functions, with particular focus on
the case of covariant POVM's, and in particular for the $SU(d)$
covariance group.

\par Let us introduce the concept of universal detector, or,
more abstractly, of universal POVM. We consider a quantum
system in a Hilbert space $\cal H$, coupled to an ancilla with Hilbert
space $\cal K$. A POVM $\{\Pi_i\}$, $\Pi_i\ge 0$ and
$\sum_i\Pi_i=I_{{\cal H}}\otimes I_{{\cal K}}$ on the Hilbert space
${\cal H}\otimes{\cal K}$ is {\em universal} for the system iff there
exists a state of the ancilla $\nu$ such that for any operator $O$ one
has
\begin{equation}
{\rm Tr} [\rho O]=
\sum_i f_i(\nu,O){\rm Tr}[(\rho\otimes\nu)\Pi_i]\,,
\label{def}
\end{equation}
where $f_i(\nu,O)$---parametrized by $\nu $ and $O$---is a suitable
function of the outcome $i$ of the measurement, and we will refer to
it as the {\em data processing} function. The detector will be called
{\em universal} when it is described by a universal POVM. In order to
give a necessary and sufficient condition for universality, we need to
introduce some notation, and the concept of {\em frame of
operators}. We will use the following symbols for bipartite pure states in ${\cal
H}\otimes{\cal K}$
\begin{equation}
|A{\rangle\!\rangle}=\sum_{n=1}^{\hbox{\scriptsize dim}{\cal H}}
\sum_{m=1}^{\hbox{\scriptsize dim}{\cal K}} 
A_{nm}|n\rangle \otimes|m\rangle \;,\label{iso} 
\end{equation}
where $|n\rangle$ and $|m\rangle$ are fixed orthonormal bases for
$\cal H$ and $\cal K$, respectively. Equation (\ref{iso})  exploits the
isomorphism \cite{bellobs} between the Hilbert space of the
Hilbert-Schmidt operators $A,B$ from ${\cal K}$ to ${\cal H}$, with
scalar product $\langle A,B\rangle =\hbox{Tr}[A^\dag B]$, and the
Hilbert space of bipartite vectors
$|A{\rangle\!\rangle},|B{\rangle\!\rangle}\in {\cal H}\otimes{\cal
K}$, with ${\langle\!\langle} A|B{\rangle\!\rangle}\equiv\langle A,B\rangle $. 
It is easy to show
the following identities
\begin{equation}
\begin{split}
&A\otimes B\dket{C}=\dket{ACB^\tau}\,,\\
&{\rm Tr}_{\cal K}[\dket{A}\dbra{B}]=AB^\dag\,,\\
&{\rm Tr}_{\cal H}[\dket{A}\dbra{B}]=A^\tau B^*\,,
\end{split}
\label{ids}
\end{equation}
where $\tau$ and $*$ denote transposition and complex conjugation 
with respect to the fixed bases. 
\par A {\em frame}\cite{czz} for operators---say $A$ from ${\cal
  K}$ to ${\cal H}$---is just a set of operators spanning a normed
linear space of operators, i. e. there are constants $0<a\le b<\infty$ 
such that for all operators $A$ one has $a|\!|A|\!|^2\le\sum_i |\langle
A,\Xi_i\rangle|^2\le b|\!|A|\!|^2$. Here, for simplicity, we will
consider the (Hilbert) space of Hilbert-Schmidt operators from ${\cal
  K}$ to ${\cal H}$, and use the equivalent vector notation introduced
in Eq. (\ref{iso}). Frames of operators have been already used
disguised as {\em spanning sets of operators}\cite{macca}  in the context of quantum
tomography. For $\{\Xi_i \}$ an operator frame there exists another
frame $\{\Theta _i \}$---called {\em dual frame}---providing operator
expansions in the form 
\begin{equation}
A=\sum _i {\rm Tr}[\Theta^\dag  _i A]\Xi _i\;.\label{sset}
\end{equation}
The completeness relation of the frame and its dual reads
\begin{equation}
\sum _i \langle \psi |\Xi _i |\phi \rangle \langle \varphi | \Theta
^\dag _i |\eta \rangle =\langle \psi |\eta \rangle \langle
\varphi|\phi \rangle \;, \label{ort}
\end{equation}
for any $\phi, \varphi \in {\cal H}$ and $\psi ,\eta \in {\cal K}$.
For continuous sets, the sums in Eqs. (\ref{sset}) and (\ref{ort}) are
replaced by integrals. Given a frame $\{\Xi_i \}$, generally
the dual set is not unique. However, all duals $\{\Theta_i\}$ of a
given frame can be obtained via the linear relation\cite{li}
\begin{eqnarray}
|\Theta_i\rangle\!\rangle=F^{-1}|\Xi_i\rangle\!\rangle+|Y_i\rangle\!\rangle-\sum_j \langle\!\langle\Xi_j|F^{-1}|\Xi_i\rangle\!\rangle|Y_j\rangle\!\rangle\,,
\;\label{duals}
\end{eqnarray}
where $Y_i$ are arbitrary, and the positive and invertible operator
$F$ writes
\begin{eqnarray}
F=\sum_i|\Xi_i\rangle\!\rangle\langle\!\langle\Xi_i|\,.
\;\label{frame}
\end{eqnarray}
The operator $F$ is called {\em frame operator} in frame theory
\cite{czz}, whereas the set of operators corresponding to the vectors
$F^{-1}|\Xi_i\rangle\!\rangle$ through the above isomorphism is called
{\em canonical dual frame}.  As we will show immediately, the dual frame
provides the data processing function, whence Eq. (\ref{duals}) allows
a useful flexibility in the data-processing, with the possibility of
optimizing the statistical error in the estimation by minimization
over the free operators $Y_i$.

\par Let us now consider a universal POVM on ${\cal H}\otimes{\cal
K}$.  The elements $\{\Pi_i\}$ can be diagonalized as follows
\begin{equation}
\Pi_i=\sum_{j=1}^{r_i}\dket{\Psi^{(i)}_j}\dbra{\Psi^{(i)}_j}\,,
\label{diagp}
\end{equation}
where the vectors $\dket{\Psi^{(i)}_j}$ have norm equal to the $j$-th
eigenvalue of $\Pi_i$, and $r_i$ is the rank of $\Pi_i$. From 
the normalization condition $\sum _i \Pi _i= I_{\cal H}\otimes I_{\cal
  K}$, it follows that the set of operators $\{\Psi^{(i)}_j\}$ from
${\cal K}$ to ${\cal H}$ must be an operator frame itself. The
characterization of universal POVM's is then given by the 
condition that there exists a state $\nu \in {\cal K}$ such that the
following operators
\begin{equation}
\Xi_i[\nu]\equiv\sum_{j=1}^{r_i}\Psi^{(i)}_j\nu^\tau\Psi^{(i)\dag}_j\label{xinu}
\end{equation}
are a frame for operators on ${\cal H}$. In fact, using
Eq. (\ref{diagp}),  Eq. (\ref{def}) rewrites
\begin{equation}
{\rm Tr} [\rho O]=\sum_i f_i(\nu,O){\rm Tr}
\left[\rho\sum_{j=1}^{r_i}\Psi^{(i)}_j\nu^\tau\Psi^{(i)\dag}_j\right]\,,
\label{CNES}
\end{equation}
and this is true independently of $\rho$ iff
\begin{equation}
O=\sum_if_i(\nu,O)\Xi_i[\nu]\,.\label{iff}
\end{equation}
From linearity one has
\begin{equation}
f_i(\nu,O)={\rm
  Tr}[\Theta^\dag_i[\nu]O] \;,\label{lin}
\end{equation}
where $\Theta_i[\nu]$ is a dual frame of $\Xi_i[\nu]$. Hence, after
  finding a dual frame for $\Xi_i[\nu]$, the data processing function is
  easily evaluated via Eq. (\ref{lin}).  

\par When restricting our attention just on the system Hilbert space,
notice that from Eqs. (\ref{xinu}) and (\ref{iff}) a universal
detector corresponds to a system POVM whose elements make a frame of
positive operators. Then, from Eqs. (\ref{iff}) and (\ref{lin}) it
follows that such POVM is ``informationally complete''\cite{Busch}, namely it 
allows evaluation of the expectation of an arbitrary system
operator. Since the number of elements of an operator frame for $\cal H$
cannot be smaller than $(\hbox{dim }{\cal H})^2$, an info-complete
POVM is necessarily not orthogonal. Viceversa, it is simple to prove
that an arbitrary frame for operators in $\cal H$ made of positive
operators $\{K_i\}$ allows to construct an info-complete POVM. In
fact, since the operator $S\equiv \sum _i K_i $ is invertible, the set $\{\tilde
K_i =S^{-1/2}K_i S^{-1/2} \}$ satisfies the completeness relation
$\sum _i \tilde K_i = I_{\cal H}$. The direct construction of
info-complete POVM's is not trivial, since it involves the searching
of {\em positive} operator frames. A way to construct universal
POVM's is suggested by group-theoretic techniques\cite{prl}. For
example, one can consider projectors on 
maximally entangled states, namely a Bell POVM on ${\cal
  H}\otimes{\cal H}$.  In the notation of Eq. (\ref{iso}), a Bell POVM
has elements of the form
\begin{equation}
\Pi_i=\frac{\alpha_i}d\dket{U_i}\dbra{U_i}\,,
\end{equation}
where $d$ is the dimension of $\cal H$, $\alpha_i$ are suitable
positive constants and $U_i$ are unitaries.  When the POVM is
orthogonal, one has $\alpha _i =1$ and $\hbox{Tr}\left[U^\dag _i U _j
\right]=d\, \delta _{ij }$. Particular cases of Bell POVM's are those
in which $U_i$ are a unitary irreducible representation (UIR) of some
group ${\mathbf G}$. As an example, consider a projective UIR of an
abelian group, which therefore satisfies the relation
\begin{equation}
U _\alpha U_\beta U _\alpha ^\dag = e^{i c(\alpha ,\beta )} U_\beta \;.
\label{cab}
\end{equation}
In this case the Bell POVM is orthogonal, with number of elements
equal to the cardinality of the group $d^2$. One can show\cite{prl}
that for any ancilla state $\nu $ such that 
$\mathrm{Tr}[U^\dag _\alpha \nu^\tau]\neq
0$ for all $\alpha $, the set of $\Xi_\alpha
[\nu]=\frac1d U_\alpha \nu^\tau U^\dag _\alpha $ is an operator frame. 
By identifying $U_1 \equiv I$, a possible choice of the ancilla
state is
\begin{equation}
\nu = \frac 1d I + \frac {1}{d (d^2-1)} \sum_{\alpha  > 1} U_\alpha \;.
\end{equation} 
The dual frame in this case is unique, and is given by 
\begin{equation}
\Theta _\alpha  [\nu]=\frac {1}{d} \sum _{\beta  =1}^{d^2}\frac{U_\beta }
{\hbox{Tr}\left[U_\beta   \nu ^* \right]}\, e^{-ic(\beta  ,\alpha )} \;.\label{dua}
\end{equation} 
Correspondingly, according to Eq. (\ref{lin}), also the data
processing function is unique.  

\par There are universal Bell POVM's also from non-abelian groups. An
interesting example is provided by the group $SU(d)$. In this case the
universality of the corresponding Bell POVM is proved by showing that
the set of $\Xi_\alpha[\nu]= U_\alpha\,\nu ^\tau\,U^\dag_\alpha$ is an
operator frame.  Let us start by evaluating the frame operator, which is
given through Eq. (\ref{frame}) by
\begin{eqnarray}
F&=&\int{\rm d}\alpha\,(U_\alpha\otimes U^*_\alpha)|\nu^\tau\rangle\!\rangle\langle\!\langle\nu^\tau |(U^\dag_\alpha\otimes U^\tau _\alpha)\nonumber\\
&=&\frac1d|I\rangle\!\rangle\langle\!\langle I|+\frac{d{\rm Tr}[(\nu^\tau)^2]-1}{d^2-1}\left(I-\frac1d|I\rangle\!\rangle\langle\!\langle I|\right)\,,
\end{eqnarray}
where we used Shur's lemma to compute the integral.  It can be noticed
that $F$ is expressed in diagonal form with eigenvalues $1$ and
$\frac{d{\rm Tr}[(\nu^\tau)^2]-1}{d^2-1}$, thus it is invertible for any
$\nu^\tau $ unless ${\rm Tr}[(\nu^\tau)^2]=d^{-1}$, 
corresponding to the state $\nu=I/d$. The expression for
the inverse of the frame operator is easily evaluated
\begin{equation}
F^{-1}=\frac1d|I\rangle\!\rangle\langle\!\langle I|+\frac{d^2-1}{d{\rm Tr}[(\nu^\tau)^2]-1}\left(I-\frac1d|I\rangle\!\rangle\langle\!\langle I|\right)\,.
\label{invfrop}
\end{equation}
The canonical dual set $\Theta^0_\alpha[\nu]$ for $\Xi_\alpha[\nu]$ is
is obtained by definition as follows
\begin{equation}
|\Theta^0_\alpha[\nu]\rangle\!\rangle=F^{-1}|U_\alpha\,\nu^\tau U_\alpha^\dag\rangle\!\rangle\,,
\end{equation}
and one has
\begin{equation}
\Theta^0_\alpha[\nu]=a\,U_\alpha\,\nu^\tau U_\alpha^\dag+b\,I\,,
\label{dualsud}
\end{equation}
where $a=\frac{d^2-1}{d{\rm Tr}[(\nu^\tau)^2]-1}$ and $b=\frac{{\rm
    Tr}[(\nu^\tau)^2]-d}{d{\rm Tr}[(\nu^\tau)^2]-1}$. According to Eq.
(\ref{lin}) the processing function corresponding to the canonical
dual frame is then
\begin{equation}
f_\alpha(\nu,O)=a\,{\rm Tr}[U_\alpha\,\nu^\tau U_\alpha^\dag O]+b{\rm Tr}[O]\,.
\end{equation}
\par The knowledge of the canonical dual frame allows to parameterize all the alternate
duals as in Eq. (\ref{duals}) by the arbitrary operators $Y_\alpha$,
greatly simplifying the task of optimizing the statistical error
in the estimate of a given operator. Such "noise" can be generally defined in
terms of the eigenvalues of the covariance matrix
\begin{equation}
C=
\begin{pmatrix}
\overline{({\rm Re}f)^2}-\overline{{\rm Re}f}^2 & \overline{{\rm Re}f{\rm Im}f}-\overline{{\rm Re}f}\,\overline{{\rm Im}f}\\
\overline{{\rm Re}f{\rm Im}f}-\overline{{\rm Re}f}\,\overline{{\rm Im}f} & \overline{({\rm Im}f)^2}-\overline{{\rm Im}f}^2
\end{pmatrix}\,,
\end{equation}
where
\begin{equation}
\overline g=\int{\rm d}\alpha\,g_\alpha(\nu,O){\rm
  Tr}[\rho\Xi_\alpha[\nu]]\,. 
\end{equation}
The noise clearly depends on the state on which the estimate is done.
For a state-independent definition of noise one could use either the
maximum noise or the average noise over all (pure or mixed) states. If
one considers averages of Hermitian operators, the imaginary parts of
the processing functions can be discarded, and this is equivalent to
consider ${\rm Re}f$ only. The noise can thus be evaluated by the customary
variance $\overline{({\rm Re}f)^2}-\overline{{\rm Re}f}^2$. As an
example, we now evaluate the optimal dual frame for the estimation of
Hermitian operators, restricting our attention on covariant dual frames,
i. e.  of the form
\begin{equation}
\Theta_\alpha[\nu]=U_\alpha\xi U^\dag_\alpha\,.
\end{equation}
It can be proved that such a set is a dual frame of $\Xi_\alpha[\nu]$ iff
\begin{equation}
{\rm Tr}[\xi]=1\,,\quad{\rm Tr}[\nu^\tau \xi]=d\,.
\end{equation}
Since we are considering Hermitian operators, the processing function
can be written
\begin{equation}
{\rm Re}f_\alpha(\nu,O)+i{\rm Im}f_\alpha(\nu,O)={\rm Tr}[U_\alpha({\rm Re}\xi)U_\alpha^\dag O]+i{\rm Tr}[U_\alpha({\rm Im}\xi)U_\alpha^\dag O]\,,
\end{equation}
where ${\rm Re}\xi=\frac12(\xi+\xi^\dag)$, and ${\rm
  Im}\xi=\frac1{2i}(\xi-\xi^\dag)$. As stated before, we can restrict
attention on ${\rm Re}f_\alpha(\nu,O)$, and thus we need to consider only
the self-adjoint case $\xi\equiv{\rm Re}\xi$. Our optimization
consists in minimizing the average variance over all pure states, namely
\begin{eqnarray}
(\Delta_\xi O^2)&=&\frac1d\int{\rm d}\beta\int{\rm d}\alpha\,\langle\psi_0|U^\dag_\beta\Theta_\alpha[\nu]U_\beta|\psi_0\rangle{\rm Tr}[\Xi_\alpha^\dag[\nu]O]^2\nonumber\\
&&\quad-\frac1d\int{\rm d}\beta\,\langle\psi_0|U^\dag_\beta OU_\beta|\psi_0\rangle^2\,,
\label{covnoise}
\end{eqnarray}
where the pure states have been parametrized as
$U_\beta|\psi_0\rangle$, for a fixed arbitrary $\psi_0\in{\cal H}$ and
$U_\beta\in SU(d)$. We will compare Eq. (\ref{covnoise}) with the variance
of the ideal measurement of $O$ averaged over all pure states, namely
\begin{equation}
(\Delta_{\rm obs}O^2)=\frac1d\int{\rm d}\alpha\,\langle\psi_0|U^\dag_\alpha O^2U_\alpha|\psi_0\rangle-\frac1d\int{\rm d}\alpha\,\langle\psi_0|U^\dag_\alpha OU_\alpha|\psi_0\rangle^2\,.
\label{optnoise}
\end{equation}
Equations (\ref{covnoise}) and (\ref{optnoise}) can be evaluated using
the following identities
\begin{eqnarray}
&&\int{\rm d}\alpha\,U_\alpha AU_\alpha^\dag={\rm Tr}[A]I\nonumber\\
&&\int{\rm d}\alpha\,U_\alpha^{\otimes2} A{U_\alpha^{\otimes2}}^\dag=\frac{2}{d+1}{\rm Tr}[P_S A]P_S+\frac{2}{d-1}{\rm Tr}[P_A A]P_A\nonumber\\
&&{\rm Tr}[E B\otimes B]={\rm Tr}[B^2]\,,
\end{eqnarray}
where $P_S=\frac12(I+E)$ and $P_A=\frac12(I-E)$ are the projections on
the totally symmetric and antisymmetric subspaces of ${\cal
  H}^{\otimes2}$, and $E$ denotes the swap operator
$E|\phi\rangle\otimes|\psi\rangle=|\psi\rangle\otimes|\phi\rangle$.  The results are
\begin{eqnarray}
(\Delta_{\rm obs}O^2)&=&\frac1{d+1}\left({\rm Tr}[O^2]-\frac1d{\rm Tr}[O]^2\right)\\
(\Delta_\xi O^2)&=&\frac{{\rm Tr}[\xi^2]-1}{d-1}(\Delta_{\rm obs}O^2).
\label{noise}
\end{eqnarray}
The optimization can be achieved by minimizing the coefficient
$\frac{{\rm Tr}[\xi^2]-1}{d-1}$ with the constraints ${\rm Tr}[\xi]=1$
and ${\rm Tr}[\nu^\tau \xi]=d$. By the method of Lagrange multipliers, one
can write the variational equation
\begin{eqnarray}
\frac\delta{\delta\langle\!\langle\xi|}\left(\frac{\langle\!\langle\xi|\xi\rangle\!\rangle-1}{d-1}-\lambda\langle\!\langle\xi|\nu^\tau \rangle\!\rangle-\mu\langle\!\langle\xi|I\rangle\!\rangle\right)=0\,,
\end{eqnarray}
which leads to the following result
\begin{equation}
\xi_{\rm opt}=\frac{d^2-1}{d{\rm Tr}[(\nu^\tau)^2]-1}\nu^\tau -\frac{d-{\rm Tr}[(\nu^\tau)^2]}{d{\rm Tr}[(\nu^\tau)^2]-1}I\,,
\end{equation}
namely, the optimal covariant dual frame is the canonical one. The
optimization can be finally completed by looking for the least noisy
ancilla state. By calculating ${\rm Tr}[\xi_{\rm opt}^2]$ and
substituting in Eq. (\ref{noise}) one obtains
\begin{equation}
(\Delta_{\rm opt}O^2)=\frac{d^2+d-1-p}{dp-1}(\Delta_{\rm obs}O^2),
\label{explnoise}
\end{equation}
where $p\equiv{\rm Tr}[(\nu^\tau)^2]$. A simple differentiation of
the expression in Eq. (\ref{explnoise}) with respect to $p$ shows that
the best choice corresponds to $p=1$, namely the minimal added noise
is achieved by an arbitrary pure ancilla state.  In this case the
expression is simplified and is equal to
\begin{equation}
(\Delta_{\rm opt}O^2)=(d+2)(\Delta_{\rm obs}O^2)\,.
\end{equation}
\section*{Acknowledgments}
This work has been cosponsored by EEC through the ATESIT project
IST-2000-29681 and by MIUR through Cofinanziamento-2002. 
P. P. and M. F. Sacchi also acknowledge support from INFM through the
project PRA-2002-CLON, and G. M. D. also acknowledges partial support
from MURI program Grant No. DAAD19-00-1-0177.

\end{document}